\documentclass[10pt]{article}
\usepackage{amsmath,amssymb,amsfonts}
\usepackage{units}
\usepackage{bbold,euler}
\usepackage{graphicx}
\usepackage{dcolumn}
\usepackage{bm,bbm,mathtools,esvect}
\usepackage{slashed}
\usepackage{esvect}
\usepackage{fullpage}
\usepackage[utf8]{inputenc}
\title{\bf THE QUANTUM HARMONIC OSCILLATOR \\ AND THE REAL HILBERT SPACE}

\author{{\sf SERGIO GIARDINO\footnote{\tt sergio.giardino@ufrgs.br}}\\
\\
\small \it Departamento de Matem\'atica Pura e Aplicada \\
\small \it Universidade Federal do Rio Grande do Sul (UFRGS)\\
\small \it Caixa Postal 15080, 91501-970  Porto Alegre RS \\
\small \it Brazil}
\begin{document}
\date{} 
\maketitle

\begin{abstract}
\noindent The harmonic oscillator is considered within generalized frameworks using complex and quaternionic numbers. The classical oscillator is considered in terms of a complex position function, and quantum oscillators are examined in terms of  complex wave functions, and in terms of quaternionic wave functions as well. Both of the quantum solutions are obtained within the real Hilbert space formalism. The results reveal the complex and quaternionic descriptions as suitable frameworks for non-stationary processes, including damped oscillations, forced oscillations, and additionally self-interacting processes that cannot be appropriately described otherwise.

\vspace{2mm}

\noindent keywords: quantum mechanics; formalism; other topics in mathematical methods in physics

\vspace{1mm}

\noindent pacs numbers: 03.65.-w; 03.65.Ca; 02.90.+p.

\end{abstract}

\section{ INTRODUCTION\label{OI}}

The quantum harmonic oscillator was  firstly studied in 1926 by Erwin Schr\"odinger, and a nice historical survey of that early research   can be found in \cite{ruschka:2020alg}. However, a century of research was not enough to exhaust the possibilities of study, and the various examples of recent usage include nuclear physics \cite{Moshinsky:1959qbh,Smirnov:1996hta,Bocko:1996mpf}, damped oscillations \cite{Blasone:2000ew,Um:2002ab,Blacker:2021nto,Markus:2022zbu,Barnett:2023zzx}, anharmonic oscillations \cite{Lawrence:2024mnj,LoChiatto:2023bam}, alternative number systems \cite{Lavoie:2010ups,Djordjevic:2000zy,Arbab:2012tqh}, quantization methods \cite{marsiglio:2009thh,castro:2013msb,Almalki:2018imu,Deguchi:2020ljd}, algebraic methods \cite{ruschka:2020alg,Friedmann:2012tr}, super-symmetry \cite{Marquette:2013tyb}, curved spaces \cite{Kuru:2024btd}, non-commutative spaces \cite{vega:2014ncs}, coupled oscillation \cite{Bender:2016vdo,bruschi:2021gen,Paul:2024rdk}, non-hermitian quantum mechanics \cite{Izadparast:2019frv}, classical and quantum relations \cite{Sen:2023vor,Sajjad:2024pee}, time in quantum mechanics \cite{Coppo:2023qgh}, phase space \cite{Plavala:2021lwq,NDolo:2023dmp} modified oscillations \cite{Sen:2021huy,Nasuda:2023ahq,Avramov:2024kpt,Ghosh:2024iqs,Fernandez:2024zlx,Saner:2024mok}, anisotropic models \cite{Chadzitaskos:2024jnc,Patra:2022kpx,Kumar:2024tsz}, time dependent systems \cite{zarro:2021ftd,Soto-Eguibar:2021cag}, the  inverted oscillator potential \cite{Yuce:2021bkt}, quantum ontology \cite{tHooft:2024eck}, field theory \cite{Chakraborty:2025nvu} string theory \cite{Pereira:2024cfa}, mathematical methods \cite{Gombar:2024dlk}, massive oscillators \cite{Jafarov:2025mbq,Volkoff:2025kki}, driven oscillation \cite{Alperin:2025vib}, coherent states \cite{Shapovalov:2022cvb}, and many others.

In the present article, one adduces a novel study concerning the harmonic oscillator, and the idea is to take benefit of the analysis of the model to deep the understanding the real Hilbert space formalism, as well as to improve the comprehension of the harmonic oscillator expressed in terms of either complex wave functions, and in terms of quaternionic wave functions as well. One recalls the real Hilbert space formalism \cite{Giardino:2018rhs,Giardino:2018lem} as an approach to originally developed within the ambit of quaternionic quantum mechanics ($\mathbbm H$QM) that enabled the solution of previous drawbacks of the anti-Hermitian approach to $\mathbbm H$QM, mainly the ill-defined classical limit ({\em cf.} Sec. 4.4 of \cite{Adler:1995qqm}) that has been applied to a wide variety of quaternionic models such as the geometric phase \cite{Giardino:2016xap}, the free particle \cite{Giardino:2017yke,Giardino:2017pqq,Giardino:2024tvp}, the theorem of Virial \cite{Giardino:2019xwm}, rectangular step potentials \cite{Giardino:2020cee,Giardino:2026buh}, the quantum  scattering \cite{Giardino:2020ztf,Hasan:2019ipt}, the harmonic oscillator \cite{Giardino:2021ofo}, angular momentum \cite{Giardino:2026jvi},  spin \cite{Giardino:2023spz}, the Dirac delta potential \cite{Giardino:2025xth}, and the generalized wave equation \cite{Giardino:2023uzp}. Relativistic theories also admit quaternionic solutions in the real Hilbert space, including the Klein-Gordon equation \cite{Giardino:2021lov,Rosa:2025git}, the Dirac equation \cite{Giardino:2021mjj}, the scalar field \cite{Giardino:2022kxk}, and the Dirac field \cite{Giardino:2022gqn}.

More recently, quaternionic models have been observed to admit self-interacting solutions 
\cite{Giardino:2024tvp,Giardino:2025bym}. The self-interaction within the quantum harmonic oscillator is the subject of the present article, as well as the physical consequences of this characteristic. One has to notice the real Hilbert space of formalism within this research as partially related to the discussion of the viability of a real quantum mechanics ($\mathbbm R$QM), where wave functions and the Hilbert space are both real. There are arguments in favor of the $\mathbbm R$QM \cite{Finkelstei:2022rqm,Chiribella:2022dgr,Zhu:2020iml,Fuchs:2022rih,Vedral:2023pij}, as well as arguments against their viability \cite{Renou:2021dvp,Chen:2021ril,Wu:2022vvi}. However, the real Hilbert space presented here does not require real wave functions, and can be considered thus an alternative that could unify these formalism within a single theory with generalized wave functions.

The research on the quaternionic quantum mechanics in the real Hilbert space involves both formal and physical aspects to be considered. From a formal perspective, it represents a consistent mathematical generalization of complex quantum mechanics, and it additionally provides a natural expression for non-stationary and self-interacting physical systems. Once mathematical consistency is established, exploring phenomenological applications is certainly a crucial direction for future research. Among these, the harmonic oscillator stands out as a particularly useful element for modeling physically sophisticated system, and providing this model is the primary goal of this article.

\section{ CLASSICAL COMPLEX OSCILLATOR \label{CCO}}

Along this section, one aims to generalize the ordinary real equation describing an one-dimensional harmonic oscillation of a particle of mass $m$ and elastic strength $\kappa$, such as
\begin{equation}\label{osc00}
 m\,\ddot x+\kappa \,x=0,
\end{equation}
where $x=x(t)$ is the time dependent position function, and the dot notation means a time derivative. All the quantities of the above equation are real, and the proposed generalization of (\ref{osc00}) comprises
\begin{equation}\label{osc01}
 m\,\ddot z+k\, z=0,
\end{equation}
where $z=z(t)$ is the complex position function, and $k$ is also a complex constant. Defining the complex quantity
\begin{equation}
 u=\frac{k}{m}\qquad\mbox{so that}\qquad u=u_0+u_1i,
\end{equation}
the solution of (\ref{osc01}) is such that
\begin{equation}\label{osc05}
 z=A\exp\big[wt\big]+B\exp\big[-wt\big],
\end{equation}
where $A$ and $B$ represent complex amplitudes, and $w$ the complex constant
\begin{equation}
w=w_0+w_1i.
\end{equation}
This solution thus comprises an oscillation associated to $w_1$ and a non-oscillating feature associated to $w_0$. Moreover, it holds also the constraint
\begin{equation}\label{osc02}
 w^2=-u\qquad \Rightarrow\qquad
 \left\{ 
\begin{array}{l}
u_0=w_1^2-w_0^2\\
u_1=-2w_0w_1.
\end{array}
 \right.
\end{equation}
Of course, the real components $u_0$ and $u_1$ of $u$ correspond to the data of the problem, and the real components of $w$ must be determined in terms of them, namely
\begin{equation}
 \left\{ 
\begin{array}{l}
w^2_0=\frac{|u|-u_0}{2}\\ \\   
w^2_1=\frac{|u|+u_0}{2}
\end{array}
 \right.
 \qquad\mbox{where}\qquad |u|=\sqrt{u_0^2+u_1^2}.
\end{equation}
Therefore, establishing
\begin{equation}
 u=|u|\exp\big[\,i\theta\,\big],
\end{equation}
after some manipulation one obtains
\begin{equation}
 w=\sqrt{|u|}\exp\left[\frac{i}{2}\big(\theta\pm\pi\big)\right],
\end{equation}
that of course confirms (\ref{osc02}). Nevertheless, the above solution does not clearly illuminate the physical character of the solution. Thus, defining the notation
\begin{equation}
 z=x+iy,
\end{equation}
and by reason of (\ref{osc01}) and (\ref{osc02}), one obtains the linear system of equations
\begin{equation}\label{osc03}
\left\{
\begin{array}{l}
\ddot x+u_0x=u_1y\\ \\
\ddot y+u_0y=-u_1 x.
\end{array}
\right.
\end{equation}
This set of equations shows that the complex equation (\ref{osc02}) represents a pair of coupled oscillators. After decoupling the equations, one obtains that
\begin{equation}\label{osc04}
\ddddot x+2u_0\ddot x+|u|^2x=0,
\end{equation}
and that the variable $y$ satisfies an identical equation. The solution of (\ref{osc04}) is immediately obtained to be equivalent to (\ref{osc05}), and thus configuring either a forced or a damped oscillation, depending on the sign of $u_0$. The understanding of the classical complex oscillator as a pair of coupled oscillations is the needed physical background to obtain the quantized version of this system in the next section.

\section{COMPLEX QUANTUM OSCILLATOR\label{QCO}}
In this section, following the complex generalization of the classical oscillation (\ref{osc00}) encoded in  (\ref{osc01}), one proposes a generalized version of Schr\"odinger equation, such as
\begin{equation}\label{osc13}
\widehat H\psi=i\hbar\frac{\partial \psi}{\partial t}
\end{equation}
where the wave function $\psi$ describes oscillation of a particle of real mass $m$, complex strength $k$,
and whose Hamiltonian operator $\widehat H$ accordingly comprises
\begin{equation}\label{osc11}
 \widehat H= -\frac{\hbar^2}{2m}\frac{\partial^2}{\partial x^2}+\frac{1}{2}kx^2+ V.
\end{equation}
The complex constant $V$, namely
\begin{equation}
V= V_0+ V_1i,
\end{equation}
is usually not considered and set to zero, but their role will be further clarified in the quaternionic case. The separation of the variables is obtained using
\begin{equation}
 \psi(x,\,t)=\varphi(x)\exp\left[-\frac{E}{\hbar}t\right],
\end{equation}
where $E$, in the same token of the autonomous particle solution \cite{Giardino:2024tvp}, comprises the complex parameter
\begin{equation}
 E=E_0+E_1i.
\end{equation}
The time independent wave equation thus immediately encompasses
\begin{equation}\label{osc06}
\left( -\frac{\hbar^2}{2m}\frac{\partial^2}{\partial x^2}+\frac{1}{2}kx^2+ V+iE\right)\varphi=0.
\end{equation}
One could try the solution of the above equation decomposing the above complex equation into a system of two real equations, following what has been done in the previous case. Although his way of solution is complicated and will not be conducted, one has to keep in mind that the complex solution comprises two  coupled oscillating systems, as likewise observed in the classical complex solution. Therefore, instead of separate the real components, one defines the complex variable,
\begin{equation}\label{osc16}
 \eta(x)=w\, x
\end{equation}
where $w$ is a complex constant, as well as the wave function
\begin{equation}\label{osc17}
\varphi(x)=H(\eta)\exp\left[-\frac{\eta^2}{2}\right]
\end{equation}
which turns (\ref{osc06}) into Hermite's equation
\begin{equation}\label{osc07}
 \frac{\partial^2H}{\partial\eta^2}-2\eta\frac{\partial H}{\partial \eta}+\lambda H=0,
\end{equation}
where the $\eta^2$ term has been eliminated after choosing,
\begin{equation}\label{osc08}
 w^4=\frac{mk}{\hbar^2}
\end{equation}
as well as
\begin{equation}\label{osc09}
 1+\frac{2m\big( V+Ei\big)}{\hbar^2w^2}=-\lambda.
\end{equation}
There are several consequences to be considered. First of all, a real and negative $\lambda$ number in the right hand side of (\ref{osc09}) imposes the solution of (\ref{osc06}) in terms of confluent hypergeometric functions, inevitably non-normalizable, and singular at $x=0$, and seemingly not physically relevant. The physically relevant solutions are obtained in terms of orthogonal Hermite's polynomials if
\begin{equation}
 \lambda=2n,
\end{equation}
where $n$ is a non-negative integer number. From (\ref{osc08}-\ref{osc09}) one obtains
\begin{equation}\label{osc10}
 \left\{ 
\begin{array}{l}
\big(V_1+E_0\big)^2=\left(n+\frac{1}{2}\right)^2\frac{\hbar^2}{m}\frac{|k|-k_0}{2}\\ \\   
\big(V_0-E_1\big)^2=\left(n+\frac{1}{2}\right)^2\frac{\hbar^2}{m}\frac{|k|+k_0}{2},
\end{array}
 \right.
 \qquad\mbox{where}\qquad |k|=\sqrt{k_0^2+k_1^2}.
\end{equation}
The usual understanding of $\mathbbm C$QM associates $E_1$ as the energy of the oscillation, but such interpretation does not hold immediately within the real Hilbert space formalism, where one has to consider the expectation values, as will be seen later in this section. Therefore, one understands the above quantities as energy parameters, and following this analysis one identifies 
$V_0$ as a background energy level parameter, as expected. It is also interesting to recognize the $E_0$ parameter as related to the attenuation of the wave function (\ref{osc17}), and consequently to non-stationary processes in the time coordinate, also to be quantized. On the other hand, if $E_0=0$, as required in the usual time stationary processes, the imaginary component $V_1$ of the background potential is quantized, and their existence also depends on the imaginary component of the elastic strength $k$. Inevitably, nonzero imaginary components of $E$ and $V$ associate to non-stationary processes, and are mutually dependent on each other.

In case of real elastic strength, where of course $k_0>0$ and $k_1=0$, the quantized energy parameters (\ref{osc10}) recover the well known quantization rule of the quantum harmonic oscillator, but also imposes $V_1+E_0=0$, a situation associated to spatial and temporal non-stationary processes if the components of the sum are not identically zero. On the other hand, a complex elastic strength $k$ implies a complex parameter $w$ in (\ref{osc08}), implying the complex character of the variable of the wave function in (\ref{osc16}), and possibly changing the expectation values.

The general case, where $V,\,E\neq 0$, non-stationary oscillations take place along the time variable, but it's worth to consider interesting particular cases. If $w$ is pure imaginary, thus $k$ is real positive according to (\ref{osc08}), generating a steady state in time, but non-normalizable because of (\ref{osc17}). On the other hand, if $w\propto e^{i\pi/4}$, one obtains a real negative $k$, what can be interpreted as the inverted harmonic oscillator, also non-normalizable because of (\ref{osc17}) but a interesting result that demonstrates the unifying capacity of this solution of the complex harmonic oscillator. 

Conversely, the normalization and the orthogonality of the wave function will be dependent on the case. If $w$ contains a nonzero real component, the wave function is always normalizable, and a pure imaginary $w$ generates non-normalizable wave functions to be considered in terms of Fourier transforms. An orthogonality relation would inevitably depend on the integral
\begin{equation}
\int_{-\infty}^{\infty}\psi_n^\dagger\psi_m dx=\Bigg(\int_{-\infty}^{\infty}H_n(wx)\,H_m(\bar wx)\exp\left[-\frac{w^2+\bar w^2}{2}x^2\right]dx\Bigg)\exp\left[-\frac{E+\bar E}{\hbar}t\right],
\end{equation}
remembering that $\bar w$ is the complex conjugate of $w$. However,  only the case of real $w$ generates an orthogonality condition, and an thus emerges the important conclusion that only normalizable stationary states admit orthogonal wave functions, but normalizable states does not necessarily generate orthogonal states. In summary, the novelties of these solutions are stationary non-normalizable states in case of imaginary $w$, and normalizable non-orthogonal and non-stationary states in case of complex $w$. Therefore, the basis of the Hilbert state comprises the stationary states wave function, and the non-stationary states will be expressed in terms of this basis. The operator algebra can further illuminate the situation.

\paragraph{Algebraic solution} The complex oscillator can easily be expressed in terms of the creation and annihilation operators $\hat a^+$ and $\hat a$, respectively
\begin{equation}\label{osc30}
 \widehat a^+=\frac{1}{\sqrt{2m\,}}\Big(\widehat p+i\sqrt{mk}x\Big),\qquad \mbox{and} \qquad \widehat a=\frac{1}{\sqrt{2 m\,}}\Big(\widehat p-i\sqrt{mk}x\Big)
\end{equation}
where $\widehat p=-i\hbar\partial/\partial x$ is the usual momentum operator. Consequently, the Hamiltonian operator (\ref{osc11}) becomes
\begin{equation}
\widehat{H}= \widehat a\, \widehat a^+-\frac{\hbar}{2}\sqrt{\frac{k}{m}},
\end{equation}
Accordingly, the commutator relation turns into
\begin{equation}\label{osc12}
\big[\widehat a,\,\widehat a^+\big]=\hbar\sqrt{\frac{k}{m}}.
\end{equation}
The right hand side of (\ref{osc12}) is a complex number, and confirms previous results of quantum commutation relations involving non-stationary processes \cite{Giardino:2025bym}. 
It is not reasonable to redefine operators in order to obtain the commutator (\ref{osc12}) equal to one, because it would hide the physical interpretation involved in the complex character of the right hand side. One also points out to the non-hermitian character of the operator, so that
\begin{equation}
 \widehat a^\dagger\neq \widehat a^+\qquad \mbox{and}\qquad \left(\widehat a^+\right)^\dagger\neq \widehat a,
\end{equation}
something also due to the complex character of $k$, and also coherent with previous results of generalized quantization processes \cite{Giardino:2025jni}.
Furthermore, one also immediately obtains the analogue generalized quantized wave function
\begin{equation}\label{osc32}
 \psi_n(x,\,t)=N_n\big(\widehat a^+\big)^n\exp\left[-\frac{\sqrt{mk}}{2\hbar}x^2-\frac{E_n}{\hbar}t\right],
\end{equation}
where $N_n$ is the normalization constant, as well as the quantized complex energy parameter
\begin{equation}\label{osc29}
 E_n=i\hbar\left(n+\frac{1}{2}\right)\sqrt{mk},
\end{equation}
that reproduces (\ref{osc10}), as expected. After the complete solution has been obtained, either in analytic as well as in algebraic terms, one has to consider the expectation values in the sequel.

\paragraph{Physical expectation values} First of all, one has to remember that, in the real Hilbert space formalism 
\cite{Giardino:2018rhs}, the expectation value associated to an operator $\widehat{\mathcal O}$ differs from the usual $\mathbbm C$QM, so that
\begin{equation}\label{osc28}
\big\langle\widehat{\mathcal O}\,\big\rangle =\frac{1}{2}\int d\bm x\left[\Psi^\dagger \widehat{\mathcal O}\Psi+\Big(\widehat{\mathcal O}\Psi\Big)^\dagger\Psi\right].
\end{equation}
Remembering $\Psi^\dagger$ to be the adjoint of the wave function $\Psi$, the operator  $\widehat{\mathcal O}$ is arbitrary, and the hermiticity is not required, a condition that is removed because (\ref{osc28}) is evaluated over real numbers, and consequently the above definition is physically more general. Defining the energy operator,
\begin{equation}
 \widehat E=i\hbar\frac{\partial}{\partial t},
\end{equation}
one immediately obtains 
\begin{equation}\label{osc31}
 \big<\, \widehat E \,\big>=E_1\exp\left[-\frac{2E_0}{\hbar}t\right].
\end{equation}
The usual result is obtained in the $E_0=0$ case, and the exponential dependence on $E_0$ shows the role of this parameter to be involved in the intensity of the motion in time, something already observed in other systems 
On the other hand, simple parity arguments involving the Hermite polynomials permit to obtain
\begin{equation}
 \big<\widehat x\big>=\big<\widehat p\big>=0.
\end{equation}
On the other hand, the energy (\ref{osc31}) is immediately recovered using (\ref{osc30}), the conservation of the energy
\begin{equation}
 \widehat E=\frac{1}{2}\Big(\widehat a^+\widehat a+\widehat a \widehat a^+\Big),
\end{equation}
and 
\begin{eqnarray}
&& \widehat a \psi_n=\sqrt{\hbar n\,\mathfrak{Re}\left[\sqrt{k/m}\right]\,}\,\psi_{n-1}\\
&& \widehat a^+ \psi_n=\sqrt{\hbar (n+1)\,\mathfrak{Re}\left[\sqrt{k/m}\right]\,}\,\psi_{n+1},\\
\end{eqnarray}
where $\mathfrak{Re}[z]$ of course means the real component of $z$. Thus, the complex quantum harmonic oscillator in the real Hilbert space description is completely understood, and one can now go to the quaternionic description.

\section{QUATERNIONIC QUANTUM OSCILLATOR I\label{QQOI}}

The generalization of the complex harmonic oscillator considered in the previous section requires a quaternionic wave equation, as well as a quaternionic wave function $\Psi$. In this situation, one entertains the wave equation
\begin{equation}\label{osc25}
	\widehat{\mathcal H}\Psi=i\hbar\frac{\partial \Psi}{\partial t}
\end{equation}
where the quaternionic wave function $\Psi$ comprises two complex components $\psi_0$ and $\psi_1$, such as
\begin{equation}\label{osc26}
	\Psi=\psi_0+\psi_1 j,
\end{equation}
where $j$ is the quaternionic imaginary unit to be further entertained in  a moment. The quaternionic Hamiltonian function is set to
\begin{equation}
	\widehat{\mathcal H}= -\frac{\hbar^2}{2m}\frac{\partial^2}{\partial x^2}+\frac{1}{2}Kx^2+ U,
\end{equation}
wherein $K$ and $U$ are the quaternionic constants
\begin{equation}
	K=K_0+K_1j,\qquad U=U_0+U_1j,
\end{equation}
and $K_0,\,K_1,\,U_0$ and $U_1$ are complex constants. One also remembers $j$ as the anti-commutative quaternionic imaginary unit, so that
\begin{equation}
	j^2=-1,\qquad \mbox{and} \qquad ij=-ji.
\end{equation}
The theoretical elements of quaternionic numbers can be found elsewhere
\cite{Ward:1997qcn,Morais:2014rqc,Ebbinghaus:1990zah}, and will not be provided here. One ought to notice a second version of the wave equation (\ref{osc25}), where the imaginary unit $i$ multiplies the right hand side of time derivative of the wave function. The non-commutativity of quaternions impose this second possibility that will be considered in the next section.
Consequently, the wave equation (\ref{osc25}) and the wave function (\ref{osc26}) thus conform to a system of two linear complex differential equations
\begin{eqnarray}
	\label{osc37} && i\hbar\frac{\partial\psi_0}{\partial t}=-\frac{\hbar^2}{2m}\nabla^2\psi_0+\left(\frac{1}{2}K_0x^2+U_0\right)\psi_0-\left(\frac{1}{2}K_1x^2+U_1\right)\psi_1^\dagger\\
	\label{osc38} && i\hbar\frac{\partial\psi_1}{\partial t}=-\frac{\hbar^2}{2m}\nabla^2\psi_1+\left(\frac{1}{2}K_0x^2+U_0\right)\psi_1+\left(\frac{1}{2}K_1x^2+U_1\right)\psi_0^\dagger.
\end{eqnarray}
One observes the coupling between the wave functions if $K_1=U_1\neq 0$, and therefore this can be considered the self-interacting oscillator. The non-self-interacting case has been considered in \cite{Giardino:2021ofo}. Following the complex case (\ref{osc16}-\ref{osc17}), one proposes the complex wave functions
\begin{eqnarray}
	\label{osc39} \psi_0(x,\,t)=A_0 \,H(\eta)\exp\left[-\frac{\eta^2}{2}\right]\exp\left[-\frac{E}{\hbar}t\right],\\
	\label{osc40} \psi_1(x,\,t)=A_1 \,H(\overline\eta)\exp\left[-\frac{\overline\eta^2}{2}\right]\exp\left[-\frac{\overline E}{\hbar}t\right],
\end{eqnarray}
where $A_0$ and $A_1$ are complex amplitudes, remembering $\overline \eta$ and $\overline E$ as the complex conjugates of the correspondent complex quantities. The linear system (\ref{osc37}-\ref{osc38}), and the wave functions (\ref{osc39}-\ref{osc40}) can be reduced to the usual Hermite equation (\ref{osc07}) after imposing constraints to eliminate the $\eta^2$ term, namely
\begin{eqnarray}
	\label{osc33} \frac{\hbar^2 w^4}{m}-K_0+K_1\frac{\overline A_1}{A_0}=0\\
	\label{osc34} \frac{\hbar^2 w^4}{m}-\overline K_0-\overline K_1\frac{A_0}{\overline A_1}=0.
\end{eqnarray}
as well as 
\begin{eqnarray}
	\label{osc35} 1+\frac{2m}{\hbar^2 w^2}\left(U_0-U_1\frac{\overline A_1}{A_0}+iE\right)=-2N_0\\
	\label{osc36} 1+\frac{2m}{\hbar^2 w^2}\left(\overline U_0+\overline U_1\frac{A_0}{\overline A_1}-iE\right)=-2N_1,
\end{eqnarray}
where $N_0$ and $N_1$ are natural numbers. One has now to  work out the constraints (\ref{osc33}-\ref{osc34}) and determine the solution parameters $E$, $w$, $A_0$ and $A_1$ in terms of the initial parameters $K$ and $U$. One observes, first of all, that $K_1=A_1=0$ recovers (\ref{osc08}) from (\ref{osc33}-\ref{osc34}), and $A_1=U_1=0$ in (\ref{osc35}-\ref{osc36}) recovers (\ref{osc09}), as expected. Nonetheless, eliminating the amplitudes $A_0$ and $A_1$, from (\ref{osc33}-\ref{osc34}) one obtains
\begin{equation}
	w^4=\frac{m}{2\hbar^2}\left(K_0+\overline K_0\pm\sqrt{\big(K_0-\overline K_0\big)^2-4|K_1|^2}\,\right),
\end{equation}
or equivalently
\begin{equation}\label{osc41}
 w^4=\frac{m}{\hbar^2}\Big(\mathfrak{Re}[K]\pm i\big|\mathfrak{Im}[K]\big|\Big),
\end{equation}
which recovers (\ref{osc08}) after $K_1=0$, as desired. Additionally, (\ref{osc41}) permits to establish a comparison to the absolute value complex result (\ref{osc08})
\begin{equation}
	|w^4|=\frac{m}{\hbar^2}|K|,
\end{equation}
where
\begin{equation}
	|K|=\sqrt{|K_0|^2+|K_1|^2},
\end{equation}
thus determining (\ref{osc41}) as a generalization of the complex case (\ref{osc08}). 
Also from (\ref{osc33}-\ref{osc34}) one obtains the relation between the complex amplitudes of the wave function,
\begin{equation}
	\overline A_1=\overline Y_0 A_0,
\end{equation}
where $Y_0$ encompasses the complex quantity
\begin{equation}
	Y_0=\frac{1}{2K_1}\left(K_0-\overline K_0\pm\sqrt{\big(K_0-\overline K_0\big)^2-4|K_1|^2}\,\right),
\end{equation}
or also
\begin{equation}
	\overline Y_0=\frac{i}{K_1}\Big(\mathfrak{Im}[K_0]\mp\big|\mathfrak{Im}[K]\big|\Big).
\end{equation}
The existence of a quaternionic amplitude thus depends on a quaternionic elastic strength $K$, whose  limit $K_1\to 0$ immediately requires the quaternionic solution back to the complex case.  In other words, one cannot use a complex $K$ and a quaternionic $U$ to obtain a particular quaternionic solution. Therefore, the wave function assumes the expression
\begin{equation}\label{osc49}
	\Psi(x,\,t)=\frac{\big(1+Y_0j\big)A_0}{\sqrt{1+|Y_0|^2}|A_0|} \,H(\eta)\exp\left[-\frac{\eta^2}{2}\right]\exp\left[-\frac{E}{\hbar}t\right].
\end{equation}
It is now possible to isolate the real components of the complex energy parameter $E$ from the quantization conditions (\ref{osc35}-\ref{osc36}), initially
\begin{eqnarray}
	\label{osc44} && E_0=\frac{\overline U_0-U_0}{2i}+\frac{2N_0+1-\big(2N_1+1\big)|Y_0|^2}{N_0+N_1+1}\; \frac{\overline U_1Y_0-U_1\overline Y_0}{4i|Y_0|^2}\\
	\label{osc45} && E_1=\frac{N_1-N_0}{N_0+N_1+1}\frac{\overline U_0+U_0}{2}-\frac{2N_0+1+\big(2N_1+1\big)|Y_0|^2}{N_0+N_1+1}\,\frac{\overline U_1Y_0+U_1\overline Y_0}{4|Y_0|^2}.
\end{eqnarray}
but also
\begin{eqnarray}
	\label{osc42}&& E_0=i\left[\left(N_0+\frac{1}{2}\right)\frac{\hbar^2 w^2}{2m}+\left(N_1+\frac{1}{2}\right)\frac{\hbar^2 \overline w^2}{2m}+U_0+\frac{1-|Y_0|^2}{2Y_0}U_1\right] \\
	\label{osc43}&& E_1=\left(2N_0+\frac{1}{2}\right)\frac{\hbar^2 w^2}{2m}-\left(N_1+\frac{1}{2}\right)\frac{\hbar^2 \overline w^2}{2m}-\frac{1+|Y_0|^2}{2Y_0}U_1.
\end{eqnarray}
The reality of the parameters enables to obtain from (\ref{osc42}-\ref{osc43}) the constraints
\begin{eqnarray}
	\label{osc46}&& \frac{\overline U_0+U_0}{2}+\frac{1-|Y_0|^2}{4|Y_0|^2}\Big(\overline U_1Y_0+U_1\overline Y_0\Big)=-\frac{\hbar^2}{m}\frac{w^2+\overline w^2}{2}\big(N_0+N_1+1\big) \\
	\label{osc47}&& \frac{1+|Y_0|^2}{4|Y_0|^2}\Big(\overline U_1Y_0-U_1\overline Y_0\Big)=\frac{\hbar^2}{m}\frac{\overline w^2- w^2}{2}\big(N_0+N_1+1\big),
\end{eqnarray}
that, back in (\ref{osc44}-\ref{osc45}), generate the final expression of the quantized energy parameters
\begin{eqnarray}
	\label{uni131} && E_0=\frac{\hbar^2}{m}\,\frac{\overline w^2- w^2}{2i}\frac{N_0+\frac{1}{2}-\Big(N_1+\frac{1}{2}\Big)|Y_0|^2}{1+|Y_0|^2}-V_1\\
	\label{uni132} && E_1=\frac{\hbar^2}{m}\,\frac{\overline w^2+ w^2}{2}\,\frac{N_0+\frac{1}{2}+\Big(N_1+\frac{1}{2}\Big)|Y_0|^2}{1-|Y_0|^2}+\frac{1+|Y_0|^2}{1-|Y_0|^2}V_0.
\end{eqnarray}
Therefore, the energy expectation value is
\begin{equation}
\label{osc48}	\big<\widehat E\big>=\left[\frac{\hbar^2}{m}\,\frac{\overline w^2+ w^2}{2}\left(\frac{N_0+N_1|Y_0|^2}{1+|Y_0|^2}+\frac{1}{2}\right)+V_0\right]\exp\left[-\frac{2E_0}{\hbar}t\right],
\end{equation}
and after choosing $N_0=N_1=N$,
\begin{equation}
	\big<\widehat E\big>=\left[\frac{\hbar^2}{m}\,\frac{\overline w^2+w^2}{2}\left(N+\frac{1}{2}\right)+V_0\right]\exp\left[-\frac{2E_0}{\hbar}t\right].
\end{equation}
Evidently, the zero point energy is obtained making $N=0$ and $t=0$, so that
\begin{equation}
	\big<\widehat E\big>=\frac{\hbar^2}{m}\,\frac{\overline w^2+w^2}{4}+V_0.
\end{equation}
There are several lessons to be learned. First of all, the quaternionic oscillator is much richer in terms of the variety of energetic states, because of the existence of two quantum numbers, $N_0$ and $N_1$. The results for the energy are also compatible to a real limit, where $K$ becomes real, as well as $w$. One also observes the $|Y_0|=1$ case, that seems to be singular because of (\ref{uni131}-\ref{uni132}), but in fact do not generate singular expectation value as (\ref{osc48}) clearly shows.

As in the complex case, the majority of the quaternionic solutions are non-stationary, and therefore does not belong to the basis set of the Hilbert space, that has to be built in terms of non-self-interacting solutions \cite{Giardino:2021ofo}. The quaternionic cases permit to obtain time stationary solutions if $E_0=0$ is obtained using either a quantized $V_1$ or also $|Y_0|=1$ and $V_1=0$. Although these are singular solutions that does not generate orthogonal sets, these are quaternionic solutions that cannot be obtained in the complex case.

\section{QUATERNIONIC QUANTUM OSCILLATOR II\label{QQOII}}

In this situation, the wave equation reads
\begin{equation}\label{osc24}
\widehat{\mathcal H}\Psi=\hbar\frac{\partial \Psi}{\partial t}i,
\end{equation}
and using the wave function (\ref{osc26}) one generates the system of differential equations
\begin{eqnarray}
\label{osc14} &&\quad i\hbar\frac{\partial\psi_0}{\partial t}=-\frac{\hbar^2}{2m}\nabla^2\psi_0+\left(\frac{1}{2}K_0x^2+U_0\right)\psi_0-\left(\frac{1}{2}K_1x^2+U_1\right)\psi_1^\dagger\\
\label{osc15} && -\,i\hbar\frac{\partial\psi_1}{\partial t}=-\frac{\hbar^2}{2m}\nabla^2\psi_1+\left(\frac{1}{2}K_0x^2+U_0\right)\psi_1+\left(\frac{1}{2}K_1x^2+U_1\right)\psi_0^\dagger.
\end{eqnarray}
The solution is very similar to the first case, but the quantization conditions (\ref{osc35}-\ref{osc36}) turns into
\begin{eqnarray}
	\label{uni139} 1+\frac{2m}{\hbar^2 w^2}\left(U_0-U_1\frac{\overline A_1}{A_0}+iE\right)=-2N_0\\
	\label{uni140} 1+\frac{2m}{\hbar^2 w^2}\left(\overline U_0+\overline U_1\frac{A_0}{\overline A_1}+iE\right)=-2N_1.
\end{eqnarray}
Moreover, the energy parameters (\ref{osc44}-\ref{osc45}) thus goes to
\begin{eqnarray}
	\label{uni141}&& E_0=\frac{N_0+N_1+1}{N_0-N_1}\frac{ U_0-\overline U_0}{2i}+\frac{N_0+\frac{1}{2}-\left(N_1+\frac{1}{2}\right)|Y_0|^2}{N_0-N_1}\frac{U_1\overline Y_0-\overline U_1Y_0}{2i|Y_0|^2}\\
	\label{uni142}&& E_1=\frac{\overline U_0+U_0}{2}+\frac{N_0+\frac{1}{2}+\left(N_1+\frac{1}{2}\right)|Y_0|^2}{N_0-N_1}\,\frac{\overline U_1Y_0+U_1\overline Y_0}{2|Y_0|^2},
\end{eqnarray}
and (\ref{osc42}-\ref{osc43}) turn into
\begin{eqnarray}
	&& E_0=i\left[\left(N_0+\frac{1}{2}\right)\frac{\hbar^2 w^2}{2m}-\left(N_1+\frac{1}{2}\right)\frac{\hbar^2 \overline w^2}{2m}-\frac{1+|Y_0|^2}{2Y_0}U_1\right]\\
	&& E_1=\left(N_0+\frac{1}{2}\right)\frac{\hbar^2 w^2}{2m}+\left(N_1+\frac{1}{2}\right)\frac{\hbar^2 \overline w^2}{2m}+U_0+\frac{1-|Y_0|^2}{2Y_0}U_1.
\end{eqnarray}
The constraints (\ref{osc46}-\ref{osc47}) of the previous case become 
\begin{eqnarray}
	&& \frac{1+|Y_0|^2}{2}\frac{\overline U_1Y_0+U_1\overline Y_0}{2|Y_0|^2}=\frac{\hbar^2}{2m}\frac{\overline w^2+w^2}{2}\big(N_0-N_1\big).\\
	&& \frac{U_0-\overline U_0}{2i}+\frac{1-|Y_0|^2}{2}\frac{U_1\overline Y_0-\overline U_1Y_0}{2i|Y_0|^2}=\frac{\hbar^2}{2m}\frac{w^2-\overline w^2}{2i}\big(N_0-N_1\big),
\end{eqnarray}
and the energy parameters thus become
\begin{eqnarray}
	&& E_0=\frac{\hbar^2}{m}\,\frac{ w^2- \overline w^2}{2i}\left(\frac{N_0-N_1|Y_0|^2}{1-|Y_0|^2}+\frac{1}{2}\right)-V_1\\
	&& E_1=\frac{\hbar^2}{m}\,\frac{ w^2+\overline w^2}{2}\left(\frac{N_0+N_1|Y_0|^2}{1+|Y_0|^2}+\frac{1}{2}\right)+V_0,
\end{eqnarray}
and the choice $N_0=N_1=N$ leads to  
\begin{eqnarray}
	&& E_0=\frac{\hbar^2}{m}\,\frac{ w^2- \overline w^2}{2i}\left(N+\frac{1}{2}\right)-V_1\\
	&& E_1=\frac{\hbar^2}{m}\,\frac{ w^2+\overline w^2}{2}\left(N+\frac{1}{2}\right)+V_0.
\end{eqnarray}
The wave functions follow (\ref{osc49}), and the expectation value goes to
\begin{equation}
	\big<\widehat E\big>=E_1\exp\left[-\frac{2E_0}{\hbar}t\right].
\end{equation}
The conclusions are similar to the previous case, where the self-interacting oscillations generate non-orthogonal wave functions, and these states present a richer energy structure  than the complex oscillation. The single difference is the singularity at the point $|Y_0|=1$, where the $E_0$ parameter diverges and this divergence cannot be eliminated within the expectation value, although this limit does holds at $N_0=N_1$.

\section{CONCLUSION\label{OCO}}

In this article one considered the generalization of the harmonic oscillator using three approaches. Initially, one considered the classical description, revealing non-stationary solutions within the complex parametrization that generalizes the real description. Moreover, the quantum harmonic oscillator were analyzed in terms of the real Hilbert space approach, demonstrating that a complex strength constant, and a complex potential background enable the generation of non-stationary solutions, that are additionally non-orthogonal. The operator algebra is also obtained to this case. The third approach concerns the quantum quaternionic harmonic oscillator, also within the real Hilbert space. This third possibility admits the existence of  a self-interacting harmonic oscillator that is absent within the complex model, but that presents a richer energetic structure in terms of non-stationary oscillations.

By way of a conceptual conclusion of these results, one observes that complex numbers are a mathematical structure well suitable for considering non-stationary, and consequently to consider either dissipative or forced interactions. This inference is supported by all the three approaches considered.

There are various interesting directions for future research, including the non-stationary complex models, as well as self-interacting quaternionic models in non-relativistic, as  well as relativistic quantum mechanics.  Applications of the presented results can be researched in terms of electronic motion present in excited molecular states, where dissipative processes are expected, as well as in the generalization of the Landau quantization of oscillations that are important for describing electronic properties of materials, and finally in studying the integer quantum Hall effect of two dimensional electron systems. Also the development of quaternionic operator algebras, and non-hermitian quantum models are exciting investigation theme for the future.

\begin{footnotesize}
\paragraph{Funding} The author gratefully thanks for the financial support by Fapergs under the grant 23/2551-0000935-8 within Edital 14/2022.

\paragraph{Data availability statement} The author declares that data sharing is not applicable to this article as no data sets were generated or analyzed during the current study.

\paragraph{Declaration of interest statement} The author declares that he has no known competing financial interests or personal relationships that
could have appeared to influence the work reported in this paper.
\end{footnotesize}
%
%
%
%

\end{document}